\documentclass[reprint,superscriptaddress,twocolumn,10pt]{revtex4-1}
\usepackage{amsmath}
\usepackage{epsfig}
\usepackage{subfigure,mathrsfs}
\usepackage{array}
\usepackage{amssymb}
\usepackage{braket}
\usepackage{float}
\usepackage{lmodern,amssymb}
\usepackage{physics}
\usepackage[dvipsnames]{xcolor}

\newcommand{\beq}[0]{\begin{equation}}
\newcommand{\eeq}[0]{\end{equation}}
\newcommand{\dbar} {\ensuremath{\,\mathchar'26\mkern-12mu d}}
\def\be{\begin{equation}}
\def\ee{\end{equation}}
\def\bea{\begin{eqnarray}}
\def\eea{\end{eqnarray}}
\usepackage{hyperref}

\begin{document}
\title{Statistical Generalization of Regenerative Bosonic and Fermionic Stirling Cycles}

\author{Nikhil Gupt}
\affiliation{Indian Institute of Technology Kanpur, 
Kanpur,Uttar Pradesh 208016, India}
\author{Srijan Bhattacharyya}
\affiliation{Indian Institute of Technology Kanpur, 
Kanpur,Uttar Pradesh 208016, India}
\author{{Arnab Ghosh}}
\email{arnab@iitk.ac.in}
\affiliation{Indian Institute of Technology Kanpur, 
Kanpur,Uttar Pradesh 208016, India}

\begin{abstract}
We have constructed a unified framework for generalizing the finite-time thermodynamic behavior of statistically distinct bosonic and fermionic Stirling cycles with regenerative characteristics. In our formalism, working fluid consisting of particles obeying Fermi-Dirac and Bose-Einstein statistics are treated under equal footing and modelled as a collection of non-interacting harmonic and fermionic oscillators. In terms of frequency and population of the two oscillators, we have provided an interesting generalization for the definitions of heat and work that are valid for classical as well as non-classical working fluids. Based on a generic setting under finite time relaxation dynamics, novel results on low and high temperature heat transfer rates are derived. Characterized by equal power, efficiency, entropy production, cycle time and coefficient of performance, thermodynamic equivalence between two types of Stirling cycles is established in the low temperature ``quantum'' regime.

\end{abstract}
	
\maketitle

\section{INTRODUCTION}

\par Finite-time thermodynamic performance of a large class of quantum engines and refrigerators has gained a lot of theoretical interest in recent times~\cite{sieniutycz1991book,udo2012stochastic,
kosloff2013quantum,kosloff2014quantum,
gelbwaser2015thermodynamics,
vinjanampathy2016quantum,
binder2019thermodynamics,
myers2020bosons,*myers2021quantum}. Quantum analog of several classical cycles, such as Otto, Carnot, Stirling, Ericsson, Brayton etc, have been introduced in this context~\cite{salamon1980minimum,salamon1981finite,rubin1982optimal,chen1991effect,geva1992classical,
geva1992quantum,chen1994maximam,chen1998effect,
sisman1999power,chen1999comprehensive,
arnaud2002carnot,bhattacharyya2001comment,geva2002irreversible,
henrich2007quantum,abah2012single,
kosloff2017quantum,thomas2019quantum}. While the efficiency of the Carnot cycle is found to be independent of the nature of the working medium, efficiencies of other quantum engines are, in general, dependent on the properties of working substance~\cite{kosloff1984quantum,chatterjee2021temperature,
gelbwaser2018single}. In particular, finite time operation of a quantum Stirling cycle in presence of a regenerator~\cite{chen1993regenerative,chen1996general,
he2002quantum,lin2003performance,
lin2003optimal,chen2002performance,
wu1998performance,kaushik2001finite}, experiences distinct relaxation dynamics for different choices of environments. Its introduction in the form of an internal heat exchanger (also known as ``economizer'' by Robert Stirling), recycles heat within hot and cold parts of the cycle and makes the machine more efficient and economical. Thus, the performance of Stirling engine and refrigerators, to a large extent depends upon the specific nature of working substance, heat baths and their interactions and thereby exhibit great diversity and huge complexity.

\par Two basic models for working fluid, namely, harmonic oscillators~\cite{lin2003performance,lin2003optimal} and spin-$\frac{1}{2}$ systems~\cite{he2002quantum,chen2002performance} are used to study regenerative quantum Stirling cycle. However, the operator algebra describing quantum harmonic oscillator is very different from that of spin-$\frac{1}{2}$ system~\cite{louisell1990book}. Spin-$\frac{1}{2}$ operators have no classical analog~\cite{geva1992quantum,geva1992classical} and follow anticommutation relations compared to harmonic oscillator operators which follow bosonic commutation relations. Since, spin-$\frac{1}{2}$ systems are fermions, they adhere to Pauli exclusion principle and Fermi-Dirac statistics as opposed to harmonic oscillator working medium complying with Bose-Einstein statistics. As a result, profound distinction between Stirling cycles comprising spin-$\frac{1}{2}$ and harmonic oscillator working mediums are quite natural~\cite{he2002quantum,chen2002performance,lin2003performance,lin2003optimal}. Keeping in view of this distinctive nature of two kinds of working substances, we present a simple unifying model for the finite time thermodynamics of quantum Stirling cycle which can treat both the spin-$\frac{1}{2}$ and harmonic oscillator working mediums on an equal footage. To capture the fermionic character of spin-$\frac{1}{2}$ system, we model this system by a fermionic analog of harmonic oscillator which is in one-to-one correspondence with Pauli spin matrices~\cite{das1993field}. This provides the major motivation for the study of regenerative Stirling cycles with fermionic and bosonic oscillators, as undertaken here.

\par Differential behavior of a fermionic oscillator in comparison with the conventional harmonic oscillator have been emphasized in multiple occasions, particularly, in the context of parametric control, quantum dissipative dynamics, dissociation of molecular dimers, just to name a few~\cite{ghosh2014parametric,ghosh2012fermionic,
ghosh2015parametric}. Nevertheless, an in-depth understanding of their thermodynamic implications in quantized settings deserves its own merit. Interestingly, we have shown throughout our work, in spite of substantial differences between two oscillators in several respects, many close thermodynamic parallels can be established between the performance of fermionic and the more familiar ones for bosonic (harmonic) Stirling cycles. Based on the unique generalization of heat and work, number of thermodynamic quantities of Stirling cycles involving classical and non-classical working fluids are computed within a uniform setup. From the general solutions of population dynamics, intriguing results on heat transfer rates are derived under near and far from equilibrium conditions. Performance characteristics of the engine and refrigerator cycles are investigated for several interesting cases with special emphasis to low temperature ``quantum'' limit.

\par Present work is organized as follows: In Sec.~\ref{Sec. II}, basic model of the Stirling cycle is introduced and general expressions for heat and work are identified. In Sec.~\ref{Sec. III}, cycle diagrams of Stirling engine and refrigerators with regenerative characteristic are discussed and amount of heat exchange and work done are obtained for harmonic and fermionic counterparts. Based on the quantum master equation and semigroup approach, time evolution of the population dynamics and heat conduction rates are calculated in Sec.~\ref{Sec. IV}. Using closed form expressions for cycle times in Sec.~\ref{Sec. V}, finite-time performance in Sec.~\ref{Sec. VI} for both cycles are shown to be statistically equivalent at low temperature. Finally, we conclude in Sec.~\ref{Sec. VII}.

\section{BASIC FORMALISM}\label{Sec. II}

\par For our study, the working fluid is modelled as a collection of noninteracting harmonic or fermionic oscillators~\cite{das1993field,ghosh2014parametric,ghosh2012fermionic,
ghosh2015parametric}. We introduce the fermionic oscillator in a similar way as the Hamiltonian of a harmonic (bosonic) oscillator  is expressed in terms of annihilation ($\hat{a}$) and creation ($\hat{a}^{\dag}$) operators 
\begin{equation}\label{H-BO}
\hat{H}_{B}=\frac{\hbar\omega}{2}(\hat{a}^{\dag}\hat{a}+\hat{a}\hat{a}^{\dag}),
\end{equation}
satisfying the commutation relation $[\hat{a},\hat{a}^{\dag}]=1$. The symmetrical structure of the Hamiltonians indicates that we are dealing with Bose particles, while for fermionic systems, an underlying asymmetry is the natural choice. Thus the Hamiltonian of a fermionic oscillator  with frequency $\omega$ is represented by
\begin{equation}\label{H-FO}
\hat{H}_{F}=\frac{\hbar\omega}{2}(\hat{a}^{\dag}\hat{a}-\hat{a}\hat{a}^{\dag}),
\end{equation}
in terms of fermionic operators, obeying $\{\hat{a}, \hat{a}^{\dag}\}$=1.

\par With the help of commutation (anticommutation) relation of the bosonic (fermionic) operators, working medium (or ``system'') Hamiltonian of Eqs.~\eqref{H-BO} and~\eqref{H-FO} can be expressed as    
\begin{equation}\label{HO-FO-Hamiltonian}
\hat{H_s} =\hbar \omega\left(\hat{a}^{\dag}\hat{a}\pm\frac{1}{2}\right)=\hbar\omega\left(\hat{N}\pm\frac{1}{2}\right).
\end{equation}
Here the plus and minus sign refers to the harmonic and fermionic oscillator respectively, while the number operator defined as $\hat{N}=\hat{a}^{\dag}\hat{a}$, satisfies the eigenvalue equation, 
\begin{eqnarray}
\hat{N}|n_{B}\rangle &=& n_{B} |n_{B}\rangle, \quad\quad n_B=0,1,2,....\infty \nonumber\\
\hat{N}|n_{F}\rangle &=& n_{F} |n_{F}\rangle, \quad\quad n_F=0,1.
\end{eqnarray}
for the respective oscillators. As a consequence, the Hilbert space of harmonic oscillator is unbounded and infinite dimensional, while that of fermionic oscillator, it is bounded and two dimensional, with operators in one-to-one correspondence with Pauli spin matrices.

\par In view of the Hamiltonian of a spin-$\frac{1}{2}$ system $H=\frac{1}{2}\hbar\omega \sigma_z$, where $\omega$ is proportional to the external magnetic field, suggests that the oscillator's frequency plays the role of an external field~\cite{geva1992quantum,chen2002performance}. With this analogy, varying the magnitude of the external magnetic field, one can change the oscillator frequency $\omega$ in time, and thereby harmonic (fermionic) Stirling engine (refrigerator) is carried out along a closed path. Note that the magnetic field can take on both negative or positive values, but in both cases the frequency of the oscillator is always positive.

\par One pertinent point to keep in mind that commutation relations obeyed by bosons having the same algebra as classical Poisson brackets does not imply fermions which do not fulfil these algebraic relations, can't have a classical limit. Fermionic number operator and the Hamiltonian operator do have classical limits because they are bilinear in $\hat{a}$, $\hat{a}^{\dag}$, and commute with each
other~\cite{ghosh2014parametric,ghosh2012fermionic,
ghosh2015parametric}. For example, working fluid obeying Fermi-Dirac statistics, quantities like work, power, heat currents, can be measured classically as they are bilinear combination of fermionic creation and annihilation operators. On the other hand, anticommutation relation obeyed by fermions are something very special appearing only in quantum mechanics. It incorporates Pauli exclusion principle which does not make sense at the classical level.

\par To this end, we define the ``temperature'' of the working system as a parameter uniquely defined by the ratio of populations between the differnt energy levels of the oscillators. For fermionic oscillator which has only two levels, this requires no further assumptions in terms of endoreversibility~\cite{callen1985book,geva1992classical}. However, for harmonic oscillator, the population ratios between different energy levels may lead to different temperatures. In both cases, the statistical average over the quantum mechanical expectation value of the number operator $\langle \hat{N} \rangle = n^{F}_{B}$ provides an useful interpretation of the endoreversibility in terms of the inverse positive ``temperature'' $\beta_s=\frac{1}{k_B T_s}$, through the relation: 
\begin{equation}\label{population}
n=\langle n^{F}_{B} \rangle_{s}=\frac{1}{\exp(\beta_s\hbar\omega)\pm1}.
\end{equation}
Here `$\pm$' sign refers to the Fermi-Dirac and Bose-Einstein distribution respectively when fermionic (harmonic) oscillators are used as the working systems of the Stirling cycle. Notice that the average occupation number of Fermi-Dirac distribution $n=\bar{n}_F$, lies between $0 \leq \bar{n}_F \leq \frac{1}{2}$, while for Bose-Einstein distribution $n=\bar{n}_{B}\geq 0$, it has only lower bound with no upper bound. This has far-reaching consequences, as we will explore shortly.

\par Following Eq.~\eqref{population}, the internal energy of the bosonic (fermionic) oscillator [Eq.~\eqref{HO-FO-Hamiltonian}] is given by 
\begin{equation}\label{internal-energy}
E=\langle \langle\hat{H_s}\rangle\rangle_{s}=\hbar\omega\left({n\pm\frac{1}{2}}\right).
\end{equation}
Immediately, one can infer that the working system may change its internal energy by changing the frequency of the oscillator or by changing its population via
\begin{equation}\label{dE-expression}
dE=\hbar\left( n\pm\frac{1}{2}\right) d\omega + \hbar\omega dn.
\end{equation}   
Comparing the above equation with the differential form of the first law of the thermodynamics
\begin{equation}\label{first-law}
dE= \dbar W + \dbar Q,
\end{equation}
we can identify the terms on the right hand side of Eq.~\eqref{dE-expression} with the inexact differential form of heat and work as
\begin{gather}
\dbar Q=\hbar\omega dn,
\label{change-in-heat}
\end{gather}	
and
\begin{gather}
	\dbar W=\hbar\left( n\pm\frac{1}{2}\right) d\omega,
\label{workdone}
\end{gather}
respectively. Eqs.~\eqref{change-in-heat} and~\eqref{workdone} are the first important result of our analysis. Throughout our paper, we follow the convention, $\dbar Q$ is positive, if heat is flowing into the system and $\dbar W$ is negative, if work is done by the system. Several remarks are now in order:

i) Since, $\bar{n}_B \geq 0$ and $0 \leq \bar{n}_F \leq \frac{1}{2}$, for positive $d\omega$, Eq.~\eqref{workdone} says that $\dbar W$ is always positive for harmonic oscillator and negative for fermionic oscillator, implying that work is done on the system for bosonic oscillator while it is done by the system for fermionic case.

ii) Since harmonic oscillator has a classical analog with frequency inversely proportional to the volume of the classical fluid~\cite{blickle2012realization}, Eq.~\eqref{workdone} for negative $d\omega$ corresponds to work done by the classical fuild in an expansion process. Such ``classical'' correspondence can't be made for fermionic oscillator for which system frequency is the only physically controllable parameter.  

iii) Equation~\eqref{dE-expression} can  therefore be regarded as the generalized version of the first of law of thermodynamics~\cite{niedenzu2018quantum} that holds good for both classical and non-classical fuild having different statistical properties.

iv) Above discussion for energy, work, and heat are calculated for single fermionic or bosonic oscillator. Hence, it is justified to multiply by the total number of noninteracting particles to get the same quantities for the working fluid as a whole.

v) As the change in the internal energy over the cycle is zero, i.e., $\oint dE =0$, we find from Eqs.~\eqref{first-law}-\eqref{workdone} that the total output work per cycle is
\begin{equation}\label{w-tot-gen}
-W_{tot}=-\hbar\oint (n \pm 1/2) d\omega = \hbar\oint \omega d n = \oint \dbar Q. 
\end{equation}
We will use Eq.~\eqref{w-tot-gen} in the following sections to calculate total output (input) work for both types of regenerative Stirling cycles. However, the present scheme can be generalized to other engine and refrigerator cycles as well. Notably, such generalization, especially in the field of quantum thermodynamic cycles, is not well-known.

\section{REGENERATIVE STIRLING CYCLE}\label{Sec. III}

\par
\begin{figure}
\centering
\includegraphics[width=\columnwidth]{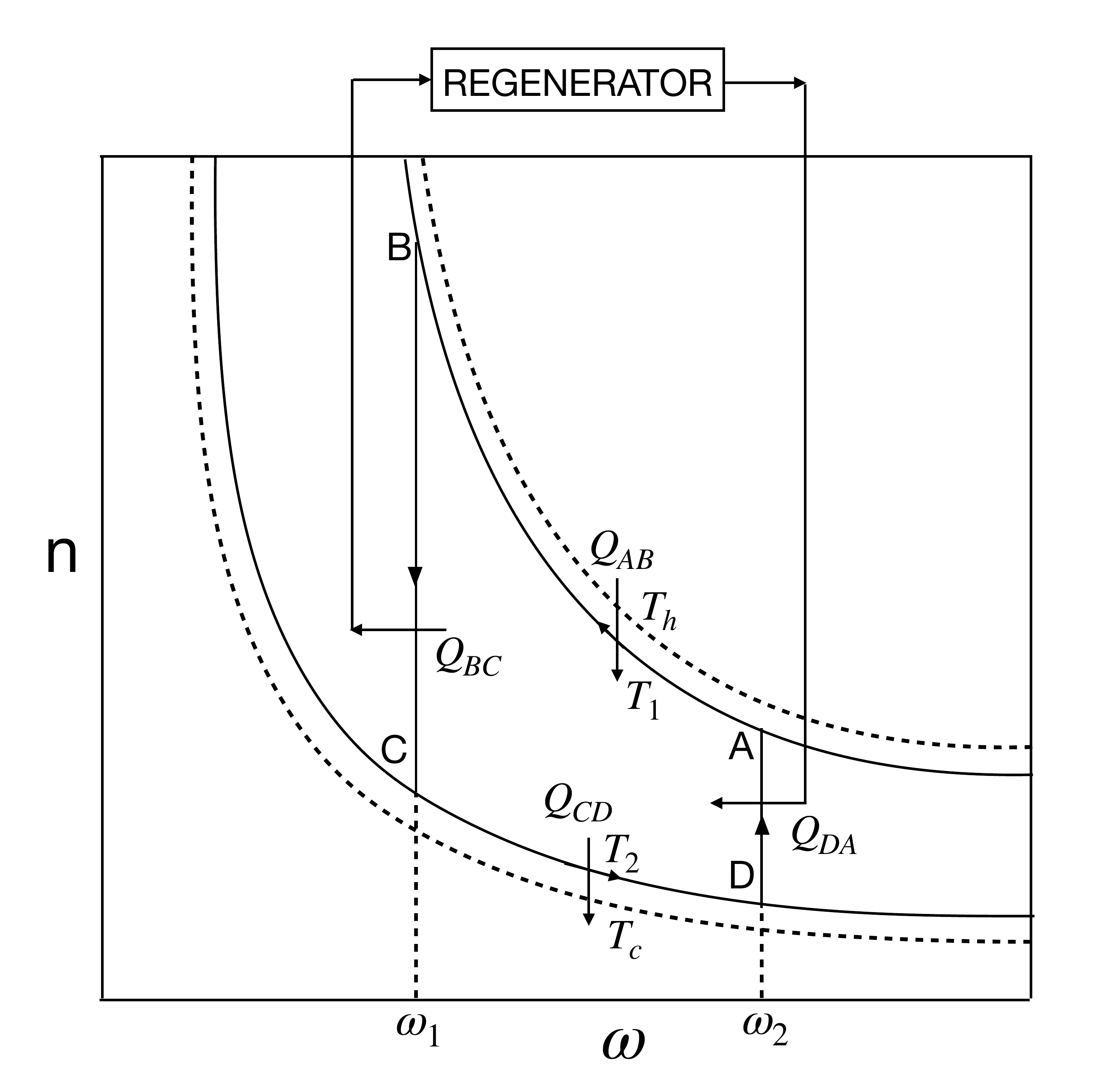}
\caption{Schematic $n-\omega$ diagram of a quantum Stirling heat engine with regenerative characteristics.}
\label{fig:1}
\end{figure}

 Cycle diagrams of quantum Stirling engine  and refrigerators have been shown in Figs.~\ref{fig:1} and~\ref{fig:2} respectively. Each figure consists of four strokes, two isothermal and two isochoric processes. The direction and the amount of heat flow along the various strokes have been shown explicitly in both the diagrams. Since the dependence of $n$ and $\omega$ is of Boltzmann type [Eq.~\eqref{population}], the isothermal branches of the cycles (AB and CD) look like exponentials in the $n-\omega$ planes, whereas the isochoric strokes are characterized by the constant frequency (BC and AD) lines at $\omega_1$ and $\omega_2$, where without loss of any generality we set $\omega_1< \omega_2$~\cite{lin2003performance,lin2003optimal,he2002quantum,chen2002performance}. Throughout our paper, the hot bath temperature is $T_h$ and the cold bath temperature is $T_c$ which is assumed to be higher than the condensation temperature of the working fluid~\cite{ghosh2017born,*ghosh2017born-kothari}.

\par Now, the direction of heat flow and work done can be understood from Eqs.~\eqref{change-in-heat} and~\eqref{workdone}. In case of engine [Fig.~\ref{fig:1}], work is negative (for bosonic case) from $A \rightarrow B$, i.e., work is done by the system along the hot isothermal branch, while it is positive for fermionic engine. Opposite is true for the cold isothermal branch ($C \rightarrow D$) for which work done is positive for harmonic and negative for fermionic counter. Keeping in view that the frequency of the harmonic oscillator is inversely proportional to the volume of the classical fluid~\cite{blickle2012realization}, this can be corroborated with the standard sign convention of work followed by any classical fluid undergoing isothermal volume expansion and compression processes. However, this is no way in contradiction with fermionic engine, once frequency is chosen as the only relevant system parameter for the non-classical fluid. So, in the following sections, we describe the isothermal expansion and compression processes only in terms of frequency change which will allow us to develop a systematic treatment for both types of working fluids in a universal way. In both cases, the heat is absorbed by the system along the hot isothermal branch and released along the cold one.

\par In case of refrigeration, the direction of heat flow gets reversed to that of engine cycle. Further, the two constant frequency processes are connected with a regenerator~\cite{chen1993regenerative,chen1996general,
he2002quantum,lin2003performance,
lin2003optimal,chen2002performance,
wu1998performance,kaushik2001finite}, located in between hot and cold segments of the engine or refrigerator. It stores heat from one cycle ($Q_{BC}$ or $Q_{AD}$) and uses it in the next cycle ($Q_{DA}$ or $Q_{CB}$).

\subsection{REGENERATIVE CHARACTERISTICS}

\subsubsection{Heat Engine}
	  
\par The net amount of heat exchange between the working system and the regenerator can be calculated by adding $Q_{BC}$ and $Q_{DA}$ i.e. $\Delta{Q}=Q_{BC} + Q_{DA}$. Now, we can have three possibilities~\cite{lin2003performance,chen2002performance}: (i) $\Delta{Q}=0$, this is the case of perfect regeneration and we have $|Q_{BC}|=|Q_{DA}|$ i.e. the amount of heat flowing from the working system to regenerator is equal to the heat flowing from regenerator to the working system. (ii) $\Delta{Q}< 0$ i.e. $|Q_{BC}|> |Q_{DA}|$, this requires that the redundant heat in the regenerator must be released in a timely manner to the cold bath. Otherwise the temperature of regenerator will change and it will not operate normally. So, the release of heat from the regenerator increases the amount of heat from $Q_{CD}$ to $Q_{CD} - |\Delta{Q}| $ to the cold bath, while heat transfer from hot bath to the system remains undisturbed. Lastly, (iii) $\Delta{Q}>0$ implies $Q_{BC}< Q_{DA}$, i.e., the inadequate heat in the regenerator must be compensated by the heat from the hot bath. This increases the amount of heat from $Q_{AB}$ to $Q_{AB} +\Delta{Q}$, by the hot bath to the working system, while the heat flowing from system to the cold bath is kept constant. As a consequence, heat released from the hot bath to the working system can be expressed in a compact form as 
\begin{gather}
Q_h= Q_{AB} + \delta{\Delta{Q}},
\label{Q_h-QHE}
\end{gather}   
where $\delta=0$ for $\Delta{Q}\leq 0$ and $\delta=1$ for $\Delta{Q}> 0$.

\subsubsection{Refrigerator}

\par In this case total heat exchange is given by $\Delta{Q}=Q_{AD}+Q_{CB}$. Similar to the engine case, here are also three possibilities~\cite{lin2003optimal,he2002quantum}: (i) $\Delta{Q}=0$, i.e. perfect regeneration. (ii) $\Delta{Q}<0$, i.e., $|Q_{CB}|< |Q_{AD}|$, so the redundant heat in the regenerator must be released to the cold bath that will decrease the net amount of heat absorption from the cold bath from $Q_{DC}$ to $Q_{DC}-\delta|\Delta{Q}|$. Finally, (iii) $\Delta{Q}>0$ for $Q_{CB}> Q_{AD}$, i.e., the inadequate heat in the regenerator per cycle must be compensated by the hot bath in timely manner. So the heat released to the hot bath reduces from $Q_{BA}$ to $Q_{BA} +\Delta{Q}$, while the heat extracted from the cold bath into the system remains unaltered. So, the net amount of heat extracted from the cold bath will be  
\begin{equation}\label{Q_c-gen}
Q_c = Q_{DC}-\delta|\Delta{Q}|,
\end{equation}  
where $\delta=0$ when $\Delta{Q} \geq 0$ and $\delta=1$ when $\Delta{Q}<0$.

\par In order to evaluate the efficiency and coefficient of performance of the Stirling cycles, in what follows, we will use Eqs.~\eqref{w-tot-gen}-\eqref{Q_c-gen} to calculate the explicit expressions for the amount of heat absorption and rejection by the system during all the processes.

\subsection{BOSONIC VS FERMIONIC ENGINE}

\par In Fig.~\ref{fig:1}, the isothermal frequency compression process from $A\rightarrow B$  occurs at system ``temperature'' $T_s=T_1$, when our system is connected with the hot bath and $C\rightarrow D$ is an isothermal compression process at ``temperature'' $T_s=T_2$ when our system is connected with the cold bath. Due to finite heat transfer rate, the ``temperatures'' $T_1$ and $T_2$ of the working system in the two isothermal processes are assumed to be different from temperatures of the heat baths and they satisfy the following relationship: $T_h>T_1>T_2>T_c$~\cite{lin2003performance,chen2002performance}. Now, using Eqs.~(\ref{population}) and~(\ref{change-in-heat}), one can go ahead and compute the general form of heat exchange (See Appendix) during all four processes of the Stirling engine as summarized below for working fluid with bosonic and fermionic statistics:	 
\begin{eqnarray}
Q_{AB}=\hbar \int_{A}^{B}\omega dn = \frac{\hbar\omega_1}{e^\frac{\hbar\omega_1}{K_BT_1}\pm 1} - \frac{\hbar\omega_2}{e^\frac{\hbar\omega_2}{K_BT_1}\pm 1} \nonumber\\
\pm {K_BT_1}\ln\left[{\frac{1\pm e^{-\frac{\hbar\omega_1}{K_BT_1}}}{1\pm e^{-\frac{\hbar\omega_2}{K_BT_1}}}}\right].
\label{heat-Q_AB}
\end{eqnarray}  
\begin{eqnarray}
Q_{CD}= \hbar\int_{C}^{D}\omega dn  &=&\frac{\hbar\omega_2}{e^\frac{\hbar\omega_2}{K_BT_2}\pm 1} - \frac{\hbar\omega_1}{e^\frac{\hbar\omega_1}{K_BT_2}\pm 1} \nonumber\\
&\pm& K_BT_2	\ln\left[{\frac{1\pm e^{-\frac{\hbar\omega_2}{K_BT_2}}}{1\pm e^{-\frac{\hbar\omega_1}{K_BT_2}}}}\right].
\end{eqnarray}      
\begin{equation}
Q_{BC}=\hbar \int_{B}^{C}\omega_1 dn = \frac{\hbar\omega_1}{e^\frac{\hbar\omega_1}{K_BT_2}\pm 1} - \frac{\hbar\omega_1}{e^\frac{\hbar\omega_1}{K_BT_1}\pm 1}.
\end{equation}  
\begin{gather}
    Q_{DA}= \hbar\int_{D}^{A}\omega_2 dn = \frac{\hbar\omega_2}{e^\frac{\hbar\omega_2}{K_BT_1}\pm 1} - \frac{\hbar\omega_2}{e^\frac{\hbar\omega_2}{K_BT_2}\pm 1}.
\label{heat-Q_DA}
\end{gather}
Total output work per cycle can then be calculated as
\begin{eqnarray}
-W_{tot} &=& Q_{AB} + Q_{BC} + Q_{CD} + Q_{DA} \nonumber\\
&=&{\pm\frac{1}{\beta_1}\ln\left(\frac{1\pm e^{-\beta_1\hbar\omega_1}}{1\pm e^{-\beta_1\hbar\omega_2}}\right) \pm \frac{1}{\beta_2}\ln\left(\frac{1\pm e^{-\beta_2\hbar\omega_2}}{1\pm e^{-\beta_2\hbar\omega_1}}\right)}.\nonumber\\
\label{total_work}
\end{eqnarray}
The value of $\Delta{Q}$ is calculated to be
\begin{eqnarray}
\Delta{Q}&=&Q_{BC}+Q_{DA}\nonumber\\
&=&\frac{\hbar\omega_1}{e^\frac{\hbar\omega_1}{K_BT_2}\pm 1} - \frac{\hbar\omega_1}{e^\frac{\hbar\omega_1}{K_BT_1}\pm 1} + \frac{\hbar\omega_2}{e^\frac{\hbar\omega_2}{K_BT_1}\pm 1} - \frac{\hbar\omega_2}{e^\frac{\hbar\omega_2}{K_BT_2}\pm 1}.\nonumber\\
\label{heat-exchange}
\end{eqnarray}
Finally, the efficiency ($\eta$) of the engine can be expressed in view of Eqs. (\ref{total_work}), (\ref{Q_h-QHE}) as     
\begin{gather}
\eta=\frac{-W_{tot}}{Q_h}=\frac{\left[ {\pm\frac{1}{\beta_1}\ln\left(\frac{1\pm e^{-\beta_1\hbar\omega_1}}{1\pm e^{-\beta_1\hbar\omega_2}}\right) \pm \frac{1}{\beta_2}\ln\left(\frac{1\pm e^{-\beta_2\hbar\omega_2}}{1\pm e^{-\beta_2\hbar\omega_1}}\right)}\right]}{Q_{AB}+\delta\Delta{Q}}.
\label{eta-QHE}
\end{gather}
	 
\subsection{BOSONIC VS FERMIONIC REFRIGERATOR}
	 
\par In case of refrigeration [Fig.~\ref{fig:2}], the process from $B\rightarrow A$ is an isothermal frequency expansion at higher temperature $T_s=T'_1$ and $D\rightarrow C$ is an isothermal frequency compression at lower temperature $T_s=T'_2$ when the system is connected with the hot and cold bath respectively. Here, finite heat transfer rate leads to the following relationship among the various temperatures involved in the entire process: $T'_1>T_h>T_c>T'_2$~\cite{lin2003optimal,he2002quantum}. In this case, the heat exchange is just the reverse to that of engine cycle. Here we summarize the expressions for heat exchange during all four processes with bosonic and fermionic fluids:
\begin{figure}
\centering
\includegraphics[width=\columnwidth]{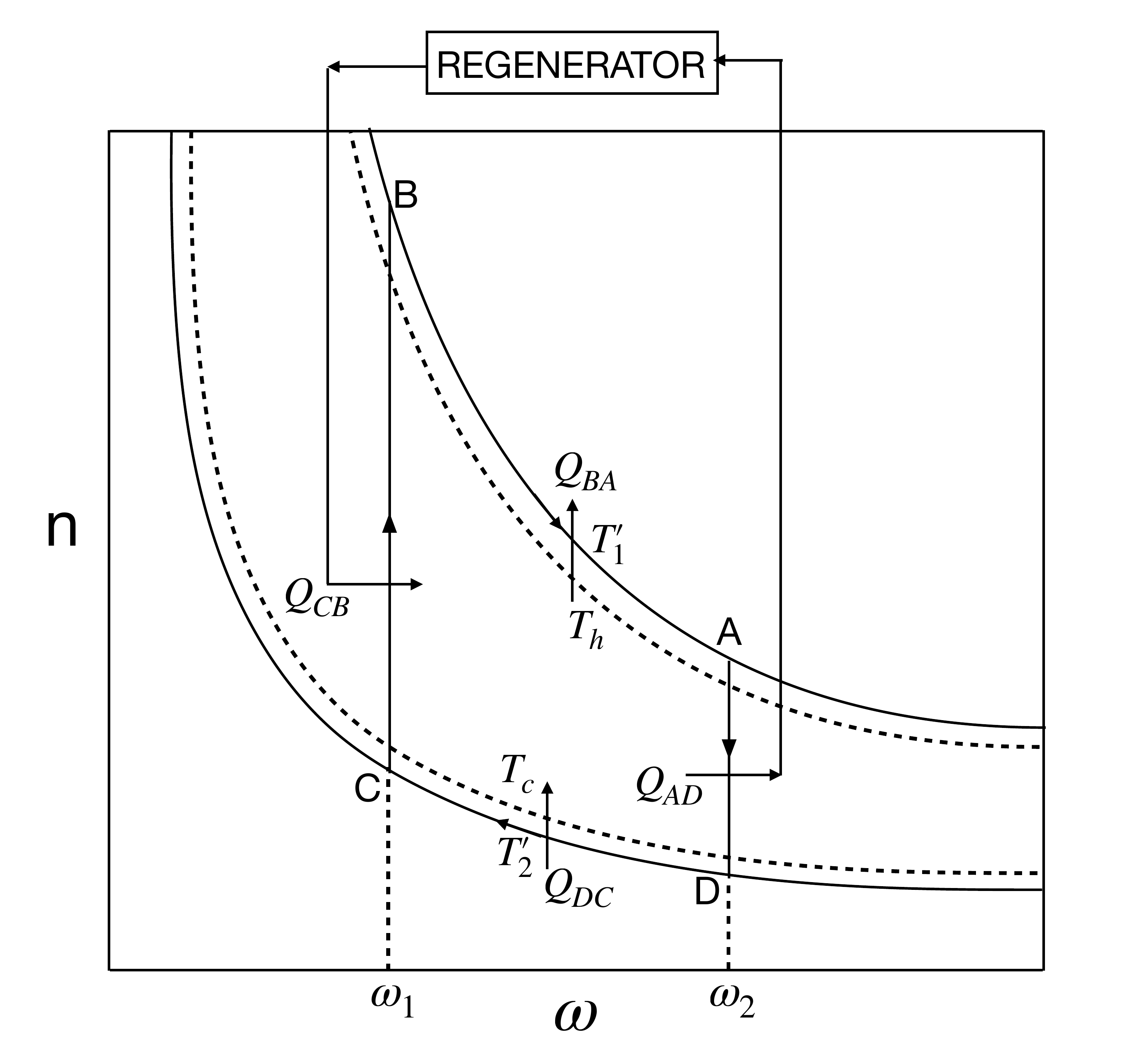}
\caption{Schematic $n-\omega$ diagram of a quantum Stirling refrigerator with regenerative characteristics.}
\label{fig:2}
\end{figure}	
\begin{equation}
Q_{BA}= \frac{\hbar\omega_2}{e^{\frac{\hbar\omega_2}{K_BT'_1}}\pm 1} - \frac{\hbar\omega_1}{e^{\frac{\hbar\omega_1}{K_BT'_1}}\pm 1}\pm K_BT'_1\ln\left[{\frac{1\pm e^{-\frac{\hbar\omega_2}{K_BT'_1}}}{1\pm e^{-\frac{\hbar\omega_1}{K_BT'_1}}}}\right].
\end{equation} 
\begin{equation}
Q_{DC}= \frac{\hbar\omega_1}{e^{\frac{\hbar\omega_1}{K_BT'_2}}\pm 1} - \frac{\hbar\omega_2}{e^{\frac{\hbar\omega_2}{K_BT'_2}}\pm 1}\pm K_BT'_2\ln\left[{\frac{1\pm e^{-\frac{\hbar\omega_1}{K_BT'_2}}}{1\pm e^{-\frac{\hbar\omega_2}{K_BT'_2}}}}\right].  
\end{equation} 
\begin{equation}
Q_{CB}= \frac{\hbar\omega_1}{e^{\frac{\hbar\omega_1}{K_BT'_1}}\pm 1} - \frac{\hbar\omega_1}{e^{\frac{\hbar\omega_1}{K_BT'_2}}\pm 1}.
\end{equation}	
\begin{equation}
Q_{AD}= \frac{\hbar\omega_2}{e^{\frac{\hbar\omega_2}{K_BT'_2}}\pm 1} - \frac{\hbar\omega_2}{e^{\frac{\hbar\omega_2}{K_BT'_1}}\pm 1}.
\end{equation} 
So, the total work done on the system by the surrounding is given by
\begin{eqnarray}
W_{tot}&=&\bigg|Q_{DC} + Q_{CB}+Q_{BA}+Q_{AD}\bigg|\nonumber\\
&=&\bigg|\pm \frac{1}{\beta'_2}\ln\left[\frac{1\pm e^{-\beta'_2\hbar\omega_1}}{1\pm e^{-\beta'_2\hbar\omega_2}}\right]\pm \frac{1}{\beta'_1}\ln\left[\frac{1\pm e^{-\beta'_1\hbar\omega_2}}{1\pm e^{-\beta'_1\hbar\omega_1}}\right]\bigg|.\nonumber\\
\label{work-QR}
\end{eqnarray}
Similarly, the value of $\Delta{Q}$ can be written as
\begin{eqnarray}
\Delta{Q}&=&Q_{AD}+Q_{CB}\nonumber\\
&=& \left[\frac{\hbar\omega_2}{e^{\frac{\hbar\omega_2}{K_BT'_2}}\pm 1} - \frac{\hbar\omega_2}{e^{\frac{\hbar\omega_2}{K_BT'_1}}\pm 1}\right] \nonumber\\
&+& \left[ \frac{\hbar\omega_1}{e^{\frac{\hbar\omega_1}{K_BT'_1}}\pm 1} - \frac{\hbar\omega_1}{e^{\frac{\hbar\omega_1}{K_BT'_2}}\pm 1}\right].
\end{eqnarray}
Now, using Eqs.~\eqref{work-QR} and~\eqref{Q_c-gen}, coefficient of performance ($\varepsilon$) for the refrigerator can be expressed as
\begin{equation}\label{rev-cop-gen}
\varepsilon=\frac{Q_c}{W_{tot}}=\frac{Q_{DC}-\delta|\Delta{Q}|}{\bigg|{\pm \frac{1}{\beta'_2}\ln\left(\frac{1\pm e^{-\beta'_2\hbar\omega_1}}{1\pm e^{-\beta'_2\hbar\omega_2}}\right)\pm \frac{1}{\beta'_1}\ln\left(\frac{1\pm e^{-\beta'_1\hbar\omega_2}}{1\pm e^{-\beta'_1\hbar\omega_1}}\right)}\bigg|}.
\end{equation}

\par We emphasize here that Eqs.~\eqref{heat-Q_AB}-\eqref{rev-cop-gen} are exact and hold for both bosonic and fermionic working mediums. If the system temperatures are equal to the heat bath temperatures, then it corresponds to reversible operation of Stirling cycle with maximum efficiency and zero power [Cf.~\eqref{eta-QHE}]. Since real engines have a finite cycle time, they cannot be in an exact equilibrium with the heat bath, consequently, their efficiency is always less than the Carnot bound~\cite{ghosh2018two-level}. Same is true for coefficient of performance of a refrigerator. Now, for finite power generation and cooling rate, dynamical laws governing the system evolution must be taken into account, where the performance of a real machine is strongly governed by heat transfer rates. So, in the next section, we first formulate the finite time dynamics~\cite{alicki1979entropy,curzon1975efficiency,
wu1999book} and then investigate heat conduction rate and machine performance at different temperature scales.

\section{FINITE TIME FORMULATION} \label{Sec. IV}

\par In order to analyze the machine performance we must solve the equation of motion that determines the time evolution of the population for both the oscillator working mediums. This is where the dynamical semigroup approach~~\cite{lindblad1976generators,
alicki1987book,breuer2002book} comes into play. It is shown that dynamical maps with semigroup properties are generated by an equation of motion with a general form 
\begin{equation}\label{eom-heisenberg}
\frac{d\hat{A}}{dt}=\frac{i}{\hbar}[\hat{H},\hat{A}]+\frac{\partial\hat{A}}{\partial{t}}+\mathscr{L}_D(\hat{A}),
\end{equation} 
where $\hat{A}$ is any system operator in the Heisenberg picture and	 
\begin{equation}
\mathscr{L}_D(\hat{A})=\sum_{\alpha}\gamma_{\alpha}(\hat{V}^\dagger_\alpha[\hat{A},\hat{V}_{\alpha}]+[\hat{V}^\dagger_\alpha,\hat{A}]\hat{V}_{\alpha}),
\end{equation}
comes from the dissipative contribution to the dynamics. $\hat{V}$, $\hat{V^\dagger}$ are system eigen-operators evaluated in their respective Hilbert spaces. $H$ is the effective system Hamiltonian and $\gamma_{\alpha}$ are the phenomenological positive damping coefficients. Equation of the form of~\eqref{eom-heisenberg} is obtained in the weak-coupling limit where the general reduction scheme staring from the microscopic Hamiltonian can be summarized as follows~\cite{breuer2002book}: 
\begin{itemize}
\item System-bath combined Hamiltonian  is considered for the dynamical evolution.  

\item  Partial trace over the bath degrees of freedom is carried out to obtain the reduced dynamical map of the
system in terms of $V(t)$.

\item  Finally, semigroup property is imposed on these reduced dynamical maps. The basic assumption is the Markovity condition: $V(t_1+t_2)=V(t_1)V(t_2)$. 
\end{itemize}

\par Now, let us consider the free Hamiltonian of the bosonic (fremionic) oscillator [Eq.~\eqref{HO-FO-Hamiltonian}] as the working system. Then we obtain $\hat{V}_{\alpha}=\hat{a},\;\hat{a}^\dagger$ as the eigen-operators of the respective oscillators satisfying commutation or anticommutation relations. Further assuming that $\hat{A}$ does not have any explicit time dependence, we obtain from Eq.~\eqref{eom-heisenberg} as:
\begin{gather}
\dot{\hat{A}}=\frac{i}{\hbar}[H_s,\hat{A}]+
\gamma_{-}(\hat{a}^\dagger[\hat{A},\hat{a}]+[\hat{a}^\dagger,\hat{A}]\hat{a})+\nonumber\\
\gamma_{+}(\hat{a}[\hat{A},\hat{a}^\dagger]+[\hat{a},\hat{A}]\hat{a}^\dagger).
\label{Heisenberg_EOM}	
\end{gather} 
Substituting $\hat{A}=\hat{N}$ into Eq.~\eqref{Heisenberg_EOM} and tracing both sides of the above equation, results in the time derivative of oscillator populations as
\begin{gather}
\dot{n}=\langle\mathscr{L}_D(\hat{N})\rangle=2\gamma_{+} \pm 2(\gamma_{-}-\gamma_{+})n.
\label{n_dot}
\end{gather}
This is second important result of our analysis which captures the population dynamics of both kind of oscillators within a single framework. Now integrating Eq.~\eqref{n_dot}, we get
\begin{equation}
\ln\left[{\frac{\mp \gamma_{+}\pm (\gamma_{-} \pm \gamma_{+})n}{\mp \gamma_{+} \pm (\gamma_{-} \pm \gamma_{+})n(0)}}\right]=-2(\gamma_{-}\pm \gamma_{+})t.
\end{equation}
The formal solution of the above differential equation can be expressed as
\begin{equation}
n=n_{eq}+(n(0)-n_{eq})e^{-2(\gamma_{-} \pm \gamma_{+})t},
\end{equation}	
where 
\begin{equation}
n_{eq}=\frac{\gamma_{+}}{\gamma_{-} \pm \gamma_{+}},
\end{equation}
is the asymptotic stationary value of $n$ which must correspond to the thermal equilibrium value of both the oscillators 
\begin{equation}
n_{eq}=\frac{1}{e^{\beta\hbar\omega} \pm 1}.
\end{equation}
Here, $\beta=\frac{1}{K_BT}$ is determined by the inverse equilibrium temperature of the reservoir (a heat bath or a regenerator) depending on the specific strokes of the cycle. Comparing above two equations we get 
\begin{gather}
\frac{\gamma_{-}}{\gamma_{+}}=e^{\beta\hbar\omega}.
\label{detailed_balence}
\end{gather}
Let us make an interesting observation.  Above equation says that $\gamma_+$ and $\gamma_-$ must satisfy Eq.~\eqref{detailed_balence} in order to ensure that the working system asmptotically reaches its correct equilibrium state, irrespective of the statistical properties of the fluid. Determination of the individual values of $\gamma_+$ and $\gamma_-$ must be based upon the reservoir correlation functions, or in other words specific models of the reservoir and the way it is coupled to the system oscillators.

One standard parametrization scheme used in the weak coupling limit when the reservoir is a thermal fields of bosonic or fermionic class, is given by~\cite{ghosh2012canonical,ghosh2012fermionic}: 
\begin{equation}\label{form-gamma-pm}
\gamma_+=\frac{\rho(\omega)}{e^{\beta\hbar\omega} \pm 1}; \quad \text{and} \quad \gamma_-=\frac{\rho(\omega)e^{\beta\hbar\omega}}{e^{\beta\hbar\omega} \pm 1}.
\end{equation}     
Here $\rho(\omega)$ is the density of states of the reservoir characterized by the coupling coefficients with the  system. The generic form of $\rho(\omega)$ satisfies a power law behavior with $\rho(\omega) \propto \omega^{m}$. For $m=1$, spectral density is ``Ohmic'', for $m>1$, it is ``super-Ohmic'' and $m<1$, it is ``sub-Ohmic''~\cite{sinha2013fluctuation,
ghosh2012canonical}. However, one must note that above parametrization scheme is true for specific bath type such as thermal radiation fields. An alternative parametrization scheme pioneered by Geva and Kosloff~\cite{geva1992quantum,geva1992classical}
\begin{gather}
\gamma_{+}=ae^{q\beta\hbar\omega}; \quad \quad \gamma_{-}=ae^{(1+q)\beta\hbar\omega},
\label{gamma_eq}
\end{gather}
where $q$ and $a$ are constant parameters determined through specific system-reservoir model, is more versatile and less restricted to a
particular reservoir type. The significance of the above simplified parametrization scheme satisfying Eq.~\eqref{detailed_balence} for all temperature ranges, is quite apparent from its widespread applications~\cite{he2002quantum,lin2003performance,
lin2003optimal,chen2002performance,
geva1992quantum,geva1992classical}. Positivity of the coefficients $\gamma_{+}, \gamma_{-}>0$ implies that the parameter $a>0$, where $a^{-1}$ defines the time scale of thermal relaxation. At high temperature, both $\gamma_+$ and $\gamma_-$ become comparable in values. On the other hand, when $\beta \rightarrow \infty$, $\gamma_+ \rightarrow 0$, and $\gamma_- \rightarrow \infty$ indicates that $q$ lies between $-1<q<0$. In what follows, we make use of Eq.~\eqref{gamma_eq} to draw a fruitful comparison between our finite time thermodynamic results valid for both low and high temperature regimes and the high temperature results derived earlier by other researchers using the same form of the transition rates~\cite{lin2003performance,
lin2003optimal,geva1992quantum}.

As a consequence, we substitute Eq.~\eqref{gamma_eq} into Eqs.~\eqref{n_dot} to obtain
\begin{gather}
\dot{n}=-2ae^{q\beta\hbar\omega}[(e^{\beta\hbar\omega} \pm 1)n-1].
\label{BOs_dn/dt}
\end{gather}
Equation~\eqref{BOs_dn/dt} plays a crucial role in our analysis. With the help of Eq.~\eqref{BOs_dn/dt}, we will calculate  the total cycle period in Sec.~\ref{Sec. V} which will be used in Sec.~\ref{Sec. VI} to analyze the finite time thermodynamic performance of Stirling cycles. Before that let us investigate the rate of heat conduction with the help of Eq.~\eqref{BOs_dn/dt} for general bosonic and fermionic systems.

\subsection{HEAT CONDUCTION RATE}
	 
\par The semigroup heat transfer rate~\cite{geva1992quantum} is defined by the time derivative of Eq.~\eqref{change-in-heat}. By substituting Eq.~\eqref{BOs_dn/dt} into Eq.~\eqref{change-in-heat}, we obtain
\begin{equation}\label{dot-Q}
\dot{Q}==-2\hbar \omega ae^{q\beta\hbar\omega}[(e^{\beta\hbar\omega} \pm 1)n-1].
\end{equation}
Here $n$ stands for the working medium populations characterized by the inverse ``temperature'' $\beta_s$ which is different from the asymptotic equilibrium temperature $\beta$ of the oscillators. Replacing $n$ by Eq.~\eqref{population} in the above equation, Eq.~\eqref{dot-Q} reduces to 
\begin{equation}\label{dot-Q-gen}
\dot{Q}==-2\hbar \omega ae^{q\beta\hbar\omega}\left[\frac{e^{\beta\hbar\omega}- e^{\beta_s\hbar\omega}}{e^{\beta_s\hbar\omega} \pm 1}\right],
\end{equation}
which is very different from any known phenomenological laws, such as linear laws of irreversible thermodynamics $\dot{Q}=L(1/T_s-1/T)$, Stephan-Boltzmann law of blackbody radiation $\dot{Q}=\alpha (T^4_s-T^4)$ or Newtonian law of heat conduction $\dot{Q}=\kappa (T_s-T)$~\cite{geva1992classical}. In all such cases, except the temperature difference, effects of all other system variables are absorbed into the phenomenological constants. Apart from being phenomenological, these laws are also derived close to thermal equilibrium ignoring complex relaxation dynamics, such as Eq.~\eqref{BOs_dn/dt} which is valid under far from equilibrium condition and depends on specific working medium statistics. From Eq.~\eqref{dot-Q-gen}, expanding $\dot{Q}$ around $\beta_s=\beta$ in first order, for two limiting cases, however, one may arrive at some form of approximate linear laws. \\

\par \textbf{High Temperature Limit:} In the high temperature limit, harmonic and fermionic systems behave differently. For harmonic oscillator, with $\beta_s\hbar\omega,\;\beta\hbar\omega \ll 1$, in Eq.~\eqref{dot-Q-gen}, we find
\begin{equation}\label{dot-Q-HO-high}
\dot{Q} \approx 2\hbar \omega a\left(\frac{\beta_s-\beta}{\beta_s}\right) = L^{high}_{HO} \Delta T. 
\end{equation}
This is precisely the same heat transfer coefficient $L^{high}_{HO}=2a k_B \hbar \omega \beta$, obtained earlier by Lin et. al~\cite{lin2003optimal,lin2003performance} for harmonic oscillator Stirling cycle (both engine or refrigerator), in the high temperature limit, which was shown to be independent of the temperature of the substance and the temperature difference between the working substance and the heat reservoir. It is evident that the Eq.~\eqref{dot-Q-HO-high} can be regarded as the Newtonian law of heat conduction and the high temperature regime can therefore be refereed as the ``classical'' limit~\cite{geva1992classical}. Under the same approximations, Eq.~\eqref{dot-Q-gen} for fermionic oscillator reduces to
\begin{equation}\label{dot-Q-FO-high}
\dot{Q}\approx a {\hbar}^{2}{\omega}^{2}(\beta_s -\beta) = L^{high}_{FO}\left(\frac{1}{T_s}-\frac{1}{T}\right).
\end{equation}
Exactly identical result with $ L^{high}_{FO}=\frac{a{\hbar}^{2}\omega^{2}}{k_B}$, was obtained earlier by Geva and Kosloff~\cite{geva1992quantum} for spin-$\frac{1}{2}$ engine at high temperature limit which is considered to be the linear law of irreversible thermodynamics.\\

\begin{figure}
\centering
\includegraphics[width=\columnwidth]{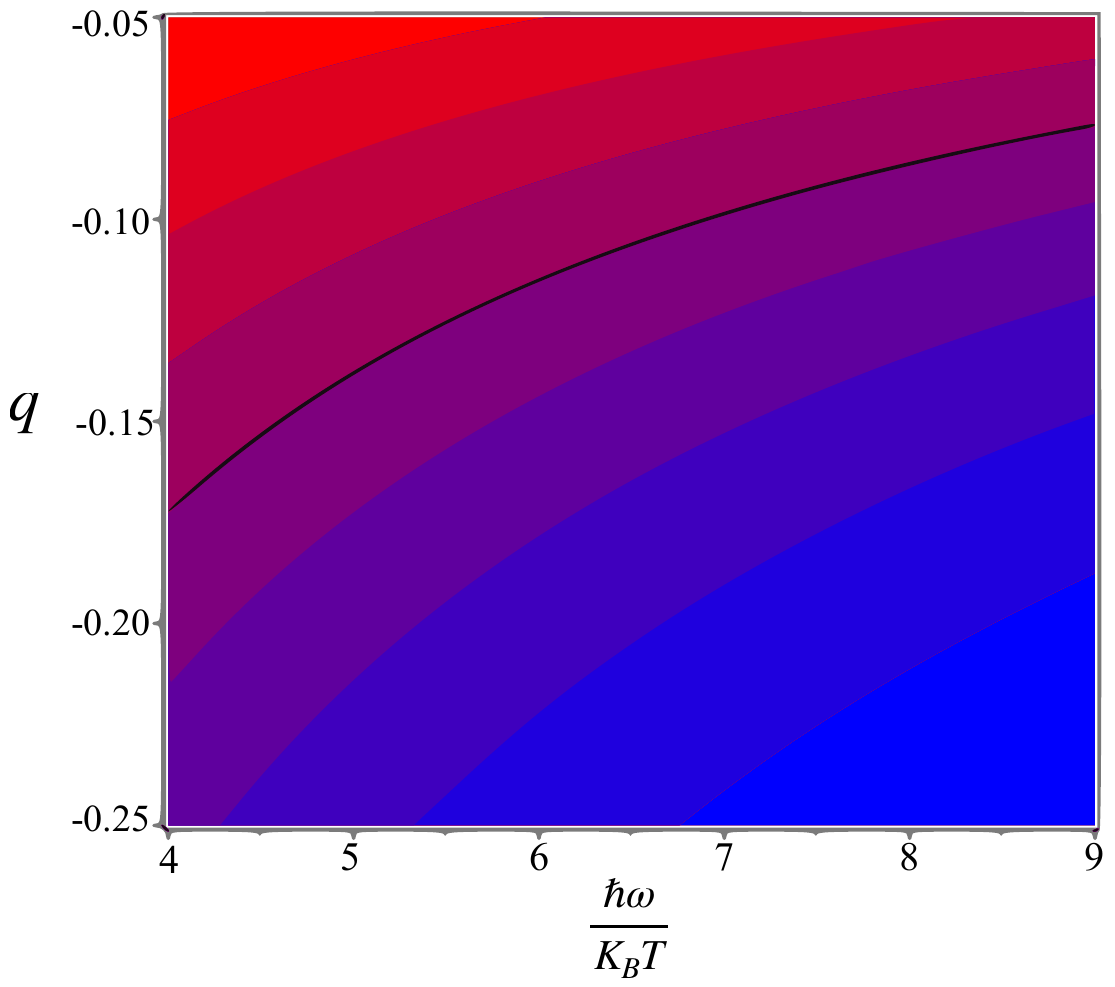}
\caption{Ratio of the heat conduction coefficients $L_r$ is plotted as a function of the parameters $q$ and $\beta \hbar \omega$. From Eq.~\eqref{L-ratio-FO}, $|q|\beta\hbar \omega = \ln 2$ line corresponds to the black curve $L_r=1$; Above the curve $L_r>1$ (red region) and below it, $L_r<1$ (blue region).}
\label{fig:3}
\end{figure}

\par \textbf{Low temperature limit:} Very low temperature limit ($\beta_s\hbar\omega, \beta\hbar\omega \gg 1$) can be referred as ``quantum''~\cite{geva1992quantum}, when we can approximate Eq.~\eqref{dot-Q-gen} as follows
\begin{equation}\label{dot-Q-low-temp}
\dot{Q}=-2\hbar \omega ae^{q\beta\hbar\omega}\left[e^{(\beta-\beta_s)\hbar\omega}- 1\right]. 
\end{equation}
Although Eq.~\eqref{dot-Q-low-temp} no longer depends on the statistical properties of working fluid, nevertheless it is not restricted to near-equilibrium situations. Further assuming, $|\beta_s -\beta|\hbar \omega \ll 1$, we recover the linear law of irreversible thermodynamics 
\begin{equation}\label{low-q-dot}
\dot{Q} \approx 2a{\hbar}^{2}\omega^{2}e^{q\beta\hbar\omega}(\beta_s -\beta) \equiv L^{low} \left(\frac{1}{T_s}-\frac{1}{T}\right). 
\end{equation}
We figure out though the heat conduction rate becomes identical for fermionic and bosonic working systems at low temperature, it strongly depends upon the bath properties via the parameter $q$. Since, $q$ is always negative, Eq.~\eqref{low-q-dot} indicates that the heat transfer rate exponentially decreases with lowering of temperature and heat exchange slows down as $q$ becomes more negative. The explicit expression for the heat transfer coefficient in this regime is given by 
\begin{equation}
L^{low}= L^{low}_{FO} \equiv L^{low}_{HO}=\frac{2a{\hbar}^{2}\omega^{2}}{k_B}e^{q\beta\hbar\omega}.
\end{equation} 
Finally, using Eqs.~\eqref{dot-Q-low-temp} and~\eqref{dot-Q-FO-high}, we can express the ratio of heat transfer coefficients of fermionic system obeying linear law of irreversible thermodynamics as:
\begin{equation}\label{L-ratio-FO}
L_r=\frac{L^{low}_{FO}}{L^{high}_{FO}}=2e^{q\beta\hbar\omega} . 
\end{equation}
Thus, depending on the values of $|q|\beta \hbar \omega \lessgtr \ln 2$, high temperature ``classical'' heat transfer coefficient of fermionic systems may be larger or smaller than the corresponding coefficient for low temperature ``quantum'' domain~[Fig.\ref{fig:3}]. This result is quite remarkable and it is a direct outcome of our present analysis.

\section{CYCLE PERIOD}\label{Sec. V}
	 
\par From Eq. (\ref{BOs_dn/dt}) we can calculate the time of heat exchange for the various strokes of the cycle. Integrating Eq.~\eqref{BOs_dn/dt} we obtain
\begin{equation}\label{gen-soln-t}
t=-\frac{1}{2a}\int^{n_f}_{n_i}\frac{dn}{e^{q\beta\hbar\omega}[(e^{\beta\hbar\omega} \pm 1)n-1]},
\end{equation}
where $n_i$ and $n_f$ are the initial and final values of $n$ along a
given path $n(\beta_{s},\omega)$ in the $n-\omega$ planes [Figs.~\ref{fig:1} and~\ref{fig:2}]. Equation~\eqref{gen-soln-t} describes the general expression for the time evolution of a harmonic (fermionic) oscillator working medium coupled with a thermal reservoir. Following Eq.~\eqref{population}, substituting the values of $n$ as a function of $\beta_{s}$ and $\omega$  and varying one parameter at a time, keeping other fixed, we can evaluate the generic formula for the time of isothermal and isochoric branches:

(a) To calculate the time of the isothermal heat exchange processes, we keep the temperature of the working system fixed at $\beta_{s}$ and vary only the frequency $\omega$. We substitute $n_i = n_i(\beta_{s}, \omega_i)$, and $n_f = n_f(\beta_{s}, \omega_f)$ into Eq.~\eqref{gen-soln-t} and obtain the following expression for an arbitrary isothermal processes as 
\begin{gather}
t=\frac{\hbar \beta_{s}}{2a}\int_{\omega_i}^{\omega_f}\frac{d\omega}{e^{q\beta\hbar\omega}(e^{\beta\hbar\omega}-e^{\beta_{s}\hbar\omega})(1\pm e^{-\beta_{s}\hbar\omega})}.
\label{isothermal_time}
\end{gather}
	
(b) In case of an isochoric process, we keep the  frequency unchanged and vary the ``temperature'' of the working substance from $\beta^{s}_{i}$ to $\beta^{s}_{f}$. Substituting $n_i = n_i(\beta^{s}_{i}, \omega)$, and $n_f = n_f(\beta^{s}_{f}, \omega)$ into Eq.~\eqref{gen-soln-t}, time of heat exchange at constant frequency (isochoric) processes can be found to be
\begin{gather}
t=\frac{\hbar\omega}{2a}\int_{\beta^{s}_{i}}^{\beta^{s}_f}\frac{d\beta_s}{e^{q\beta\hbar\omega}(e^{\beta\hbar\omega}-e^{\beta_s\hbar\omega})(1\pm e^{-\beta_s\hbar\omega})}.
\label{isochoric_time}
\end{gather}

\par Few important points are to be noted here:

(i) For isothermal process, the fixed temperature $\beta$ corresponds to the equilibrium temperature of the working system determined by the inverse heat bath temperatures (cold and hot). On the contrary, it represents the regenerator temperatures for the isochoric processes.

(ii) Equations~\eqref{isothermal_time} and~\eqref{isochoric_time} are true for both refrigeration and engine cycles. Next we will separately calculate the explicit expressions for the time involved in various strokes of the Stirling engine and refrigerators.

\subsection{In case of Stirling engine}

\par Let us go back to Fig.~\ref{fig:1} and calculate the explicit expressions for the time of different individual steps:

$\bullet$ \textbf{Stroke $A$ to $B$:} Setting $\beta=\beta_h$ and $\beta_s=\beta_1$ into Eq.~\eqref{isothermal_time}, the time required for heat exchange due to this isothermal process is found to be	 
\begin{gather}
t_1=\frac{\hbar \beta_1}{2a}\int_{\omega_1}^{\omega_2}\frac{d\omega}{e^{q\beta_h\hbar\omega}(e^{\beta_1\hbar\omega}-e^{\beta_h\hbar\omega})(1\pm e^{-\beta_1\hbar\omega})}.
\label{time_t1}
\end{gather}

$\bullet$ \textbf{Stroke $C$ to $D$:} Substituting $\beta=\beta_c$ and $\beta_s=\beta_2$ into Eq.~\eqref{isothermal_time}, we obtain the time required for heat exchange for this isothermal process as
\begin{equation}\label{time_t3}
t_3=\frac{\hbar \beta_2 }{2a}\int_{\omega_1}^{\omega_2}\frac{d\omega}{e^{q\beta_c\hbar\omega}(e^{\beta_c\hbar\omega}-e^{\beta_2\hbar\omega})(1\pm e^{-\beta_2\hbar\omega})}.
\end{equation}

$\bullet$ \textbf{Stroke $B$ to $C$:} For this constant frequency process, we put $\beta=\beta_{1r}$ and $\omega=\omega_1$ into Eq.~\eqref{isochoric_time}, then the time required to complete the stroke takes the form of
\begin{equation}\label{time_t2}
t_2=\frac{\hbar\omega_1}{2a}\int_{\beta_1}^{\beta_2}\frac{d\beta_s}{e^{q\beta_{1r}\hbar\omega_1}(e^{\beta_{1r}\hbar\omega_1}-e^{\beta_s\hbar\omega_1})(1\pm e^{-\beta_s\hbar\omega_1})},
\end{equation}
where $\beta_{1r}$ is the inverse temperature of the regenerator when heat is released from the working system to the regenerator at constant frequency $\omega=\omega_1$, so that $\beta_{1r}>\beta_s$.

$\bullet$ \textbf{Stroke $D$ to $A$:} Finally, setting $\beta=\beta_{2r}$ and $\omega=\omega_2$ into Eq.~\eqref{isochoric_time}, the time for this constant frequency process reduces to 
\begin{gather}
t_4=\frac{\hbar\omega_2}{2a}\int_{\beta_1}^{\beta_2}\frac{d\beta_s}{e^{q\beta_{2r}\hbar\omega_2}(e^{\beta_s\hbar\omega_2}-e^{\beta_{2r}\hbar\omega_2})(1\pm e^{-\beta_s\hbar\omega_2})}.
\label{time_t4}
\end{gather}
Here $\beta_{2r}$ is the inverse temperature of the regenerator when heat is transferred from the regenerator to the working system at constant frequency $\omega=\omega_2$, implying that $\beta_{2r}<\beta_s$.

\par As a result, the total cycle period is calculated to be
\begin{gather}
\tau_E =t_1+t_2+t_3+t_4.
\label{tau_QHE}
\end{gather}
It is clear from Eqs.~\eqref{time_t1}-\eqref{time_t4} that cycle periods will be different for bosonic and fermionic engines in general. Evaluating integrals of Eqs.~\eqref{time_t1}-\eqref{time_t4} in closed form for the general case is a formidable task. A closed form analytical solution can be found only in the high and low-temperature limits. Before we turn our attention to those solutions, we would like to make some general remarks about the integrands in Eqs.~\eqref{time_t1}-\eqref{time_t4}. The presence of statistical $(1 \pm e^{-x})$ factors in the denominator of Eqs.~\eqref{time_t1}-\eqref{time_t4}, are responsible for the differential behavior between the fermionic and harmonic Stirling cycles. This difference disappear in the low temperature ``quantum'' domain as the contribution from exponential terms become negligibly small. As a result both the oscillators become statistically equivalent in this temperature regime. This has profound implications on the thermodynamic performance of both the cycles. Situation gets dramatically different beyond this ``quantum'' regime. Although it is possible to obtain some closed form solutions even at the high temperature limit, nature of the explicit expressions differ significantly between the two oscillator systems.

\par Now, Let us first evaluate the explicit expressions at the low temperature limit, which will be used in Sec.~\ref{Sec. VI}. Since $-1<q<0$, if we consider Eq.~\eqref{time_t1} for example, we can approximate it at low temperature as follows:
\begin{eqnarray}
t_1 &\simeq& \frac{\hbar \beta_1}{2a}\int_{\omega_1}^{\omega_2}e^{-(\beta_1+\beta_hq)\hbar\omega}d\omega\nonumber\\
&=&\frac{\beta_1}{2a(\beta_1+\beta_hq)}(e^{-(\beta_1+\beta_hq)\hbar\omega_1}-e^{-(\beta_1+\beta_hq)\hbar\omega_2})\nonumber\\
&=&\frac{1}{2a(1+\alpha_hq)}(e^{-(1+\alpha_hq)\beta_1\hbar\omega_1}-e^{-(1+\alpha_hq)\beta_1\hbar\omega_2}),\nonumber\\
\end{eqnarray} 
where we have used the fact that $\alpha_h= \frac{\beta_h}{\beta_1}=\frac{T_1}{T_h} < 1$. Similarly, we can approximate Eq.~\eqref{time_t3} as
\begin{eqnarray}
t_3 &\simeq& \frac{\hbar \beta_2}{2a}\int_{\omega_1}^{\omega_2}e^{-\beta_c(1+q)\hbar\omega}d\omega\nonumber\\
&=&\frac{\beta_2}{2a(1+q)\beta_c}(e^{-\beta_c(1+q)\hbar\omega_1}-e^{-\beta_c(1+q)\hbar\omega_2})\nonumber\\
&=&\frac{1}{2a(1+q)\alpha_c}(e^{-\alpha_c(1+q)\beta_2\hbar\omega_1}-e^{-\alpha_c(1+q)\beta_2\hbar\omega_2}),\nonumber\\
\end{eqnarray}
where we have used the parameter $\alpha_c= \frac{\beta_c}{\beta_2}=\frac{T_2}{T_c}> 1$.

\par Calculation of $t_2$ and $t_4$ are little involved. First, we approximate Eqs.~\eqref{time_t2} and~\eqref{time_t4} as
\begin{gather}
t_2\simeq\frac{\hbar\omega_1}{2a}\int_{\beta_1}^{\beta_2}e^{-\beta_{1r}(1+q)\hbar\omega_1}d\beta_s,
\label{time-t2}
\end{gather}
and   
\begin{gather}
t_4\simeq\frac{\hbar\omega_2}{2a}\int_{\beta_1}^{\beta_2}e^{-(\beta_s+\beta_{2r}q)\hbar\omega_2}d\beta_s.
\label{time-t4}
\end{gather}
In order to solve Eq. (\ref{time-t2}) and (\ref{time-t4}), we need an extra assumption. Let us assume $\beta_{1r}\propto\beta_s$ and $\beta_{2r}\propto\beta_s$ that means $\beta_{1r} (\beta_{2r})$ and $\beta_s$ are linearly dependent with proportionality constants $\gamma_1>1$ and $\gamma_2<1$, respectively [Cf. Eqs.~\eqref{time_t2}-\eqref{time_t4}]. With this assumption, Eqs.~\eqref{time-t2} and~\eqref{time-t4} can be simplified as follows:
\begin{equation}
t_2=\frac{\hbar\omega_1}{2a\gamma_1(1+q)\hbar\omega_1}(e^{-\gamma_1(1+q)\beta_1\hbar\omega_1}-e^{-\gamma_1(1+q)\beta_2\hbar\omega_1}),
\end{equation}
and
\begin{equation}
t_4=\frac{\hbar\omega_2}{2a(1+\gamma_2q)\hbar\omega_2}(e^{-(1+\gamma_2q)\beta_1\hbar\omega_2}-e^{-(1+\gamma_2q)\beta_2\hbar\omega_2}).
\end{equation}


\par Thus one can approximate the low temperature expression for the engine cycle period as
\begin{eqnarray}\label{tau-E-final}
\tau^{low}_E&=&\frac{1}{2a(1+\alpha_hq)}(e^{-(1+\alpha_hq)\beta_1\hbar\omega_1}-e^{-(1+\alpha_hq)\beta_1\hbar\omega_2})\nonumber\\
     	&+&\frac{1}{2a(1+q)\alpha_c}(e^{-\alpha_c(1+q)\beta_2\hbar\omega_1}-e^{-\alpha_c(1+q)\beta_2\hbar\omega_2})\nonumber\\
     	&+&\frac{1}{2a(1+q)\gamma_1}(e^{-\gamma_1(1+q)\beta_1\hbar\omega_1}-e^{-\gamma_1(1+q)\beta_2\hbar\omega_1})\nonumber\\
     	&+&\frac{1}{2a(1+\gamma_2q)}(e^{-(1+\gamma_2q)\beta_1\hbar\omega_2}-e^{-(1+\gamma_2q)\beta_2\hbar\omega_2}).\nonumber\\
\end{eqnarray}
Notice that Eq.~\eqref{tau-E-final} is same for both the oscillators at low temperature region which is characteristically different from the high temperature results, derived by others~\cite{chen2002performance,he2002quantum,lin2003optimal,lin2003performance}
\begin{eqnarray}
t_1 &=& \frac{\beta_1}{4a (\beta_1-\beta_h)} \ln (\frac{\omega_2}{\omega_1}),\\
t_3 &=& \frac{\beta_2}{4a (\beta_c-\beta_2)} \ln (\frac{\omega_2}{\omega_1}),\\
t_2 &=& \frac{1}{4a} \int^{\beta_2}_{\beta_1} \frac{d \beta^{\prime}}{ \beta_{1r}-\beta^{\prime}},\\
t_4 &=& \frac{1}{4a} \int^{\beta_2}_{\beta_1} \frac{d \beta^{\prime}}{\beta^{\prime}-\beta_{2r}},
\end{eqnarray}
for spin-$\frac{1}{2}$ fermions and  
\begin{eqnarray}
t_1 &=& \frac{\omega_2-\omega_1}{2a \hbar \omega_2 \omega_1 (\beta_1-\beta_h)},\\
t_3 &=& \frac{\omega_2-\omega_1}{2a \hbar \omega_2 \omega_1 (\beta_c-\beta_2)},\\
t_2 &=& \frac{1}{2a \hbar \omega_1} \int^{\beta_2}_{\beta_1} \frac{d \beta^{\prime}}{\beta^{\prime} (\beta_{1r}-\beta^{\prime})},\\
t_4 &=& \frac{1}{2a \hbar \omega_2} \int^{\beta_2}_{\beta_1} \frac{d \beta^{\prime}}{\beta^{\prime} (\beta^{\prime}-\beta_{2r})},
\end{eqnarray}
for harmonic oscillator working mediums. One can immediately understand that at high temperature, cycle period for both the oscillators are very distinct in nature, whereas at low temperature, they become exactly identical. Secondly, we find the time required for the heat exchange at very low temperature depends heavily on the specific model of the reservoir through the $q$ parameter [Eq.~\eqref{tau-E-final}], whereas it is independent of the bath characteristics at high temperature range. As an upshot of this consequence, finite time thermodynamic behavior of bosonic and fermionic engines are shown to be equivalent in the low temperature regime, while they deviate largely in the opposite limit. Analogous situation is obtained also for refrigeration. So, in essence, we expect that these two oscillator working mediums behave in an identical fashion only at low temperature region, and beyond this, they perform very differently from each other. For completeness, we briefly mention the cycle period for refrigeration at low temperature limit, which will be helpful to analyze the machine performance in Sec.~\ref{Sec. VI}.

\subsection{In case of refrigeration cycle}
     
Like heat engine, one can compute the time for various strokes of the refrigerator [See Fig.~\ref{fig:2}]. For $\beta=\beta_c$ and $\beta_s=\beta'_2$ in Eq.~\eqref{isothermal_time}, the time required for the heat exchange at constant temperature $T^{\prime}_2$ is given by
\begin{gather}
t'_1=\frac{\hbar \beta'_2}{2a}\int_{\omega_2}^{\omega_1}\frac{d\omega}{e^{q\beta_c\hbar\omega}(e^{\beta_c\hbar\omega}-e^{\beta'_2\hbar\omega})(1\pm e^{-\beta'_2\hbar\omega})}.
\label{t1_QR}
\end{gather} 
Analogously, for $\beta=\beta_h$ and $\beta_s=\beta'_1$ in Eq.~\eqref{isothermal_time}, time of the isothermal process from $B$ to $A$ at temperature $T^{\prime}_1$ is found to be
\begin{equation}\label{t3_QR}
t'_3=\frac{\hbar \beta'_1}{2a}\int_{\omega_1}^{\omega_2}\frac{d\omega}{e^{q\beta_h\hbar\omega}(e^{\beta_h\hbar\omega}-e^{\beta'_1\hbar\omega})(1\pm e^{-\beta'_1\hbar\omega})}
\end{equation} 
Exactly in the same way, as we have done before, for $\beta=\beta'_{1r}$ in Eq. (\ref{isochoric_time}), time of heat exchange from $A$ to $D$ at constant frequency $\omega=\omega_2$ reduces to    
\begin{equation}\label{t4_QR}
t'_4=\frac{\hbar\omega_2}{2a}\int_{\beta'_1}^{\beta'_2}\frac{d\beta_s}{e^{q\beta'_{1r}\hbar\omega_2}(e^{\beta'_{1r}\hbar\omega_2}-e^{\beta_s\hbar\omega_2})(1\pm e^{-\beta_s\hbar\omega_2})}.
\end{equation}
Here $\beta'_{1r}$ is the inverse temperature of the regenerator and $\beta'_{1r}>\beta'$ because heat is flowing from the working system to the regenerator. Finally, putting $\beta=\beta'_{2r}$ into Eq.~\eqref{isochoric_time}, time of heat exchange from $C$ to $B$ at constant frequency $\omega=\omega_1$ is calculated to be   
\begin{gather}
t'_2=\frac{\hbar\omega_1}{2a}\int_{\beta'_2}^{\beta'_1}\frac{d\beta_s}{e^{q\beta'_{2r}\hbar\omega_1}(e^{\beta'_{2r}\hbar\omega_1}-e^{\beta_s\hbar\omega_1})(1\pm e^{-\beta_s\hbar\omega_1})},
\label{t2_QR}
\end{gather}
where $\beta'_{2r}$ is also the inverse temperature of the regenerator with $\beta'_{2r}<\beta'$, as heat is flowing from the regenerator to the working system in this case. 
So, we find the total cycle time as   
\begin{gather}
\tau_R=t'_1+t'_2+t'_3+t'_4
\label{tau_QR}
\end{gather}

Similar to engine, in the low temperature limit, Eqs.~\eqref{t1_QR}-\eqref{t3_QR} are reduced to the following forms 
\begin{eqnarray}
t'_1&=&\frac{\hbar \beta'_2}{2a}\int_{\omega_1}^{\omega_2}{e^{-(\beta'_2+q\beta_c)\hbar\omega}}d\omega\nonumber\\
&=&\frac{\beta'_2}{2a(\beta'_2+q\beta_c)}(e^{-(\beta'_2+q\beta_c)\hbar\omega_1}-e^{-(\beta'_2+q\beta_c)\hbar\omega_2}),\nonumber\\
\end{eqnarray}
and           
\begin{eqnarray}
t'_3&=&\frac{\hbar \beta'_1}{2a}\int_{\omega_1}^{\omega_2}{e^{-(1+q)\beta_h\hbar\omega}}d\omega\nonumber\\
&=&\frac{\beta'_1}{2a(1+q)\beta_h}(e^{-(1+q)\beta_h\hbar\omega_1}-e^{-(1+q)\beta_h\hbar\omega_2}).
\end{eqnarray}  
For Eqs.~\eqref{t4_QR} and~\eqref{t2_QR}, we proceed exactly in the same way as we did earlier. First, we approximate Eqs.~\eqref{t4_QR} and~\eqref{t2_QR} as
\begin{gather}
t'_2=\frac{\hbar \omega_1}{2a}\int_{\beta'_1}^{\beta'_2}e^{-(\beta_s+q\beta'_{2r})\hbar\omega_1} d\beta_s,
\label{approx_t'2}
\end{gather}
and      
\begin{gather}
t'_4=\frac{\hbar \omega_2}{2a}\int_{\beta'_1}^{\beta'_2}e^{-(1+q)\beta'_{1r}\hbar\omega_2} d\beta_s.
\label{approx_t'4}
\end{gather}
Next, we assume $\beta'_{1r}$ and $\beta'_{2r}$ are linearly dependent i.e $\beta'_{1r}=b\beta_s$ and $\beta'_{2r}=b'\beta_s$ where $b$, $b'$ are proportionality constants. From Eqs. (\ref{approx_t'2}) and (\ref{approx_t'4}), we get      
\begin{equation}
t'_2=\frac{1}{2a(1+qb')}(e^{-(1+qb')\beta'_1\hbar\omega_1}-e^{-(1+qb')\beta'_2\hbar\omega_1}),
\end{equation}     
and
\begin{equation}
t'_4=\frac{1}{2a(1+q)b}(e^{-(1+q)b\beta'_1\hbar\omega_2}-e^{-(1+q)b\beta'_2\hbar\omega_2}).
\end{equation}
Thus we calculate the following form of the total cycle period at low temperature 
\begin{gather}
     \tau^{low}_R=\frac{1}{2a(1+q\alpha'_c)}(e^{-(1+q\alpha'_c)\beta'_2\hbar\omega_1}-e^{-(1+q\alpha'_c)\beta'_2\hbar\omega_2})\nonumber\\
     +\frac{1}{2a(1+q)\alpha'_h}(e^{-(1+q)\alpha'_h\beta'_1\hbar\omega_1}-e^{-(1+q)\alpha'_h\beta'_1\hbar\omega_2})\nonumber\\
     +\frac{1}{2a(1+qb')}(e^{-(1+qb')\beta'_1\hbar\omega_1}-e^{-(1+qb')\beta'_2\hbar\omega_1})\nonumber\\
     +\frac{1}{2a(1+q)b}(e^{-(1+q)b\beta'_1\hbar\omega_2}-e^{-(1+q)b\beta'_2\hbar\omega_2}).
\end{gather}
where $\alpha'_h=\frac{\beta_h}{\beta'_1}>1$ and $\alpha'_c=\frac{\beta_c}{\beta'_2}<1$.
We will use this equation in Sec.~\ref{Sec. VI}.


\section{LOW TEMPERATURE EQUIVALENCE OF BOSONIC AND FERMIONIC STIRLING CYCLES} \label{Sec. VI}

\subsection{Bosonic vs Fermionic Engine}

\par The efficiency ($\eta$) and power output ($P$) are two important quantities to analyze the performance of an engine~\cite{ghosh2017catalysis}. Using Eqs.~\eqref{total_work},~\eqref{Q_h-QHE}~and~\eqref{tau_QHE}, $\eta$ and $P$ can be expressed as   
\begin{gather}
\eta=\frac{-W_{tot}}{Q_h}=\frac{\left[ {\pm\frac{1}{\beta_1}\ln\left(\frac{1\pm e^{-\beta_1\hbar\omega_1}}{1\pm e^{-\beta_1\hbar\omega_2}}\right) \pm \frac{1}{\beta_2}\ln\left(\frac{1\pm e^{-\beta_2\hbar\omega_2}}{1\pm e^{-\beta_2\hbar\omega_1}}\right)}\right]}{Q_{AB}+\delta\Delta{Q}},
\label{eta_QHE}
\end{gather}
and 
\begin{gather}
P=\frac{-W_{tot}}{\tau_E}=\frac{\left[ {\pm\frac{1}{\beta_1}\ln\left(\frac{1\pm e^{-\beta_1\hbar\omega_1}}{1\pm e^{-\beta_1\hbar\omega_2}}\right) \pm \frac{1}{\beta_2}\ln\left(\frac{1\pm e^{-\beta_2\hbar\omega_2}}{1\pm e^{-\beta_2\hbar\omega_1}}\right)}\right]}{t_1+t_2+t_3+t_4}.
\label{power_QHE}
\end{gather}
Now we can have three possibilities: (i) low temperature or ``quantum'' limit  ($\beta_\alpha\hbar\omega_{i}>>1$ or $\hbar\omega_{i}>>K_{B}T_{\alpha}$; $\alpha=1,2$ and $ i=1,2$); (ii) Intermediate regime ($\beta_1\hbar\omega_{i}<<1$ and $\beta_2\hbar\omega_{i}>>1$); and (iii) High temperature or ``classical'' limit ($\beta_\alpha\hbar\omega_{i}<<1$ or $\hbar\omega_{i}<<K_{B}T_{\alpha}$; $\alpha=1,2$ and $ i=1,2$). Fourth possibility is simply unphysical since it implies that temperature of the hot branch is lower than the temperature of cold one. Characterized by the energy and temperature scales of the working materials as we mentioned earlier, close resemblances between two kinds of engines are expected at low temperature range. Whereas they operate distinctively once they are away from this temperature scale. Secondly, finite time thermodynamic analysis in the intermediate regime becomes a non-trivial task since it requires the evaluation of complicated integrals for the cycle period which can be done only by numerical means. Therefore, we focus our attention only to low temperature ``quantum'' regime where both types of engines exhibit close thermodynamic kinship to each other.

\begin{figure}
\centering
\includegraphics[width=\columnwidth]{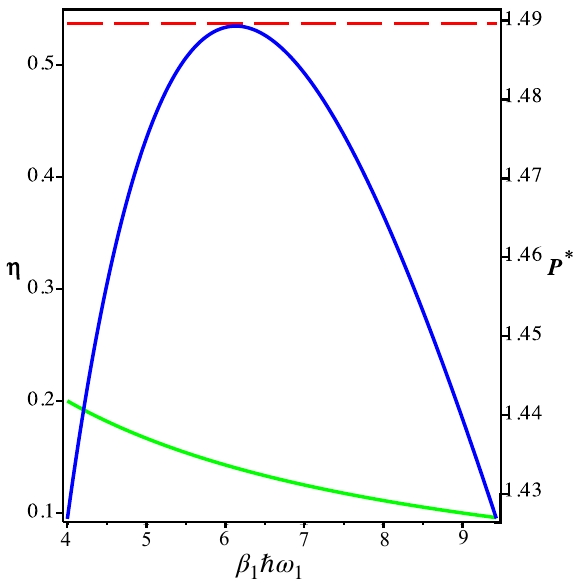}
\caption{Efficiency $\eta$ (green) and dimensionless power $P^{*}=\frac{P}{aK T_1}$ (blue) are plotted as a function of $\beta_1 \hbar \omega_1$ for the parameter set: $\omega_2=2\omega_1$, $\beta_2=2\beta_1$ ($T_1=2T_2$), $\alpha_h=\gamma_2=0.6$, $\alpha_c=\gamma_1=1.4$, $q=-0.05$. Red dotted curve represented by Curzon-Albhorn bound sets the highest possible efficiency at maximum power, where $\eta \approx 1/(1+\beta_1 \hbar \omega_1)$ [Cf.~\eqref{eta-low}] for the choice of parameters.}
	\label{fig:4}
\end{figure}
\par We have explored that low temperature thermodynamic equivalence of bosonic and fermionic Stirling engines is an upshot of the fact that all the physical quantities possess identical expressions in this temperature range, i.e., both engines attain the same expressions for heat, work, power, efficiency and even the same entropy production rate ($\sigma$) as given below: 
\begin{equation}
 \Delta{Q}=\hbar\omega_1(e^{-\beta_2\hbar\omega_1}-e^{-\beta_1\hbar\omega_1})+\hbar\omega_2(e^{-\beta_1\hbar\omega_2}-e^{-\beta_2\hbar\omega_2}),\label{delta-Q-BOs}
\end{equation}
\begin{equation}
-W_{tot}=\frac{1}{\beta_1}(e^{-\beta_1\hbar\omega_1}-e^{-\beta_1\hbar\omega_2})+\frac{1}{\beta_2}(e^{-\beta_2\hbar\omega_2}-e^{-\beta_2\hbar\omega_1}),
\end{equation}    
\begin{equation}\label{Q_h-eng-low}
 Q_h=\left(\hbar\omega_1+\frac{1}{\beta_1}\right)e^{-\beta_1\hbar\omega_1}-\left(\hbar\omega_2+\frac{1}{\beta_1}\right)e^{-\beta_1\hbar\omega_2},
\end{equation}
\begin{equation}\label{Q_c-eng-low}
Q_c =\hbar\omega_2e^{-\beta_1\hbar\omega_2}-\hbar\omega_1e^{-\beta_1\hbar\omega_1}+\frac{1}{\beta_2}(e^{-\beta_2\hbar\omega_2}-e^{-\beta_2\hbar\omega_1}),
\end{equation}
\begin{equation}
    \eta=\frac{\frac{1}{\beta_1}(e^{-\beta_1\hbar\omega_1}-e^{-\beta_1\hbar\omega_2})+\frac{1}{\beta_2}(e^{-\beta_2\hbar\omega_2}-e^{-\beta_2\hbar\omega_1})}{\left(\hbar\omega_1+\frac{1}{\beta_1}\right)e^{-\beta_1\hbar\omega_1}-\left(\hbar\omega_2+\frac{1}{\beta_1}\right)e^{-\beta_1\hbar\omega_2}}, \label{eta-low}
\end{equation}
\begin{equation}
P=\frac{\frac{1}{\beta_1}(e^{-\beta_1\hbar\omega_1}-e^{-\beta_1\hbar\omega_2})+\frac{1}{\beta_2}(e^{-\beta_2\hbar\omega_2}-e^{-\beta_2\hbar\omega_1})}{\tau^{low}_E},
     \label{power-low}
\end{equation}
\begin{gather}
\sigma=\frac{\Delta{S}}{\tau^{low}_E}=-\frac{\beta_h{Q_h}+\beta_c{Q_c}}{\tau^{low}_E},
   \label{rate_entopy_production}
\end{gather}
where $\Delta{S}$ is the entropy production determined by Eqs.~\eqref{Q_h-eng-low} and~\eqref{Q_c-eng-low}.

As a consequence, not only their reversible operations are identical, their finite time thermodynamic performances also become equivalent.   This is the most interesting observation of our analysis. We stress the reason behind the thermodynamic equivalence between the bosonic and fermionic Stirling cycles at low temperature is statistical in origin and it is different in status from the thermodynamic equivalence observed between the various engine cycles at small action limit~\cite{uzdin2015equivalence}. In the present case, origin of the equivalent performance can be attributed to the remarkable similarities between each and every expressions of harmonic engine to that of the fermionic engine. In the latter case, heat and work are significantly different within the cycle strokes, although they become equivalent over a full cycle period as a consequence of symmetric rearrangement theorem~\cite{uzdin2015equivalence}.

\par From Eqs.~\eqref{eta-low} and~\eqref{power-low}, we can see efficiency and power of both the engines strongly depend upon $q$, i.e., on the specific system-reservoir model. This behavior is truely contrasting in respect to high temperature engine performance which has no $q$ dependence. As the temperature is high enough, equipartition theorem holds and the efficiency of bosonic engine reduces to classical Carnot bound with perfect regenerative characteristics~\cite{lin2003performance}: 
\begin{eqnarray}
\Delta{Q}&=&Q_{BC}+Q_{DA}=0,\\
-W_{tot}&=&\frac{(\beta_2-\beta_1)}{\beta_1\beta_2}\ln\left(\frac{\omega_2}{\omega_1}\right),\\
Q_h&=&\frac{1}{\beta_1}\ln\left(\frac{\omega_2}{\omega_1}\right)>0,\\
Q_c&=&\frac{1}{\beta_2}\ln\left(\frac{\omega_1}{\omega_2}\right)<0,\\
\eta&=&1-\frac{\beta_1}{\beta_2}\\
P&=&\frac{-W_{tot}}{\tau^{high}_E},\\
\text{where},\nonumber\\
\tau^{high}_{HO}&=&d\left(\frac{1}{\beta_1-\beta_h}+\frac{1}{\beta_c-\beta_2}\right)+\gamma\left(\frac{1}{\beta_1}-\frac{1}{\beta_2}\right),\nonumber\\ 
\end{eqnarray}
with $d=\frac{\omega_2-\omega_1}{2a\hbar\omega_1\omega_2}$ and $\gamma$ as the proportionality constants. On the contrary, without any ``classical'' correspondence, efficiency of fermionic engine does not approach to classical Carnot bound. Yet it may work as an engine in this high temperature regime determined by the following set of quantities~\cite{chen2002performance}:
\begin{eqnarray}
\Delta{Q}&=&\frac{\hbar^2(\omega^{2}_1-\omega^{2}_2)(\beta_1-\beta_2)}{4}>0,\\
Q_h&=&\frac{\hbar^2(\omega^{2}_2-\omega^{2}_1)(\beta_1+2\beta_2)}{8}>0,\\
Q_c&=&-\frac{3\beta_2}{8}\hbar^2(\omega^{2}_2-\omega^{2}_1)<0,\\
-W_{tot}&=&\frac{\hbar^2(\omega^{2}_2-\omega^{2}_1)(\beta_2-\beta_1)}{8},\\
\eta&=&\frac{\beta_2-\beta_1}{2\beta_2-\beta_1},\\
P&=&\frac{-W_{tot}}{\tau^{high}_{FO}},\\
\text{where},\nonumber\\
\tau^{high}_{FO}&=& \frac{\beta_1}{4a (\beta_1-\beta_h)} \ln (\frac{\omega_2}{\omega_1})+\frac{\beta_2}{4a (\beta_c-\beta_2)} \ln (\frac{\omega_2}{\omega_1})\nonumber\\
&+&\gamma(\beta_2 - \beta_1).
\end{eqnarray}
Since the power output strongly depends on the $q$ parameter at low temperature scale, while it is independent in the classical limit, plays a significant role on the power maximization of the engine. It is found that Curzon-Ahlborn bound always holds in the high-temperature limit~\cite{geva1992classical,curzon1975efficiency}, irrespective of the details of the model and interestingly efficiency at maximum power for both harmonic and spin-$\frac{1}{2}$ engines are shown to abide by Carzon-Albhron bound in this temperature range. Presence of the $q$-factor, on the other hand makes the generic optimization scheme impossible at low temperature scale. Yet, for a given choice of system-reservoir interaction, engine efficiency at maximum power is found to be always less than the Curzon-Alhborn and asymptotically approaches the maximum possible value with the increase of temperature. [Fig.~\ref{fig:4}].

\subsection{Bosonic vs Fermionic Refrigerator}

Using Eqs.~\eqref{work-QR},~\eqref{Q_c-gen} and~\eqref{tau_QR}, we can obtain the expressions for the coefficient of performance ($\varepsilon$), power input ($P$) and cooling rate ($R$) as 
\begin{equation}
\varepsilon =\frac{Q_c}{W_{tot}}=\frac{Q_{DC}-\delta|\Delta{Q}|}{\bigg|{\pm \frac{1}{\beta'_2}\ln\left(\frac{1\pm e^{-\beta'_2\hbar\omega_1}}{1\pm e^{-\beta'_2\hbar\omega_2}}\right)\pm \frac{1}{\beta'_1}\ln\left(\frac{1\pm e^{-\beta'_1\hbar\omega_2}}{1\pm e^{-\beta'_1\hbar\omega_1}}\right)}\bigg|},
\end{equation} 
\begin{equation}
P =\frac{W_{tot}}{\tau_R}=\frac{\bigg|{\pm \frac{1}{\beta'_2}\ln\left(\frac{1\pm e^{-\beta'_2\hbar\omega_1}}{1\pm e^{-\beta'_2\hbar\omega_2}}\right)\pm \frac{1}{\beta'_1}\ln\left(\frac{1\pm e^{-\beta'_1\hbar\omega_2}}{1\pm e^{-\beta'_1\hbar\omega_1}}\right)}\bigg|}{t'_1+t'_2+t'_3+t'_4},
\end{equation}
\begin{equation}
R = \frac{Q_c}{\tau_R}=\frac{Q_{DC}-\delta|\Delta{Q}|}{t'_1+t'_2+t'_3+t'_4}.
\end{equation}
Similar to the engine, we can have three different regions of operation, but we will analyze only the low temperature regime, which is the most interesting regime for refrigeration as well. Like engine, refrigeration mode also possesses identical expressions between various quantities of bosonic and fermionic Stirling cycles:  
\begin{equation}
      \Delta{Q}=\hbar\omega_1(e^{-\beta'_1\hbar\omega_1}-e^{-\beta'_2\hbar\omega_1})+\hbar\omega_2(e^{-\beta'_2\hbar\omega_2}-e^{-\beta'_1\hbar\omega_2}),
\end{equation}       
\begin{equation}
     Q_c=\hbar\omega_1e^{-\beta'_2\hbar\omega_1}-\hbar\omega_2e^{-\beta'_2\hbar\omega_2}+\frac{1}{\beta'_2}(e^{-\beta'_2\hbar\omega_1}-e^{-\beta'_2\hbar\omega_2}),
\end{equation}     
\begin{equation}
W_{tot}=\bigg|\frac{1}{\beta'_2}(e^{-\beta'_2\hbar\omega_1}-e^{-\beta'_2\hbar\omega_2})+\frac{1}{\beta'_1}(e^{-\beta'_1\hbar\omega_2}-e^{-\beta'_1\hbar\omega_1})\bigg|,
\end{equation}     
\begin{equation}
\varepsilon=\frac{Q_c}{W_{tot}}=\frac{(\beta'_2\hbar\omega_1+1)e^{-\beta'_2\hbar\omega_1}-(\beta'_2\hbar\omega_2+1)e^{-\beta'_2\hbar\omega_2}}{\frac{\beta'_2}{\beta'_1}(e^{-\beta'_1\hbar\omega_1}-e^{-\beta'_1\hbar\omega_2})+(e^{-\beta'_2\hbar\omega_2}-e^{-\beta'_2\hbar\omega_1})},
\end{equation}     
\begin{equation}
P=\frac{\frac{1}{\beta'_1}(e^{-\beta'_1\hbar\omega_1}-e^{-\beta'_1\hbar\omega_2})+\frac{1}{\beta'_2}(e^{-\beta'_2\hbar\omega_2}-e^{-\beta'_2\hbar\omega_1})}{\tau^{low}_R},
\end{equation}    
and    
\begin{equation}
R=\frac{\hbar\omega_1e^{-\beta'_2\hbar\omega_1}-\hbar\omega_2e^{-\beta'_2\hbar\omega_2}+\frac{1}{\beta'_2}(e^{-\beta'_2\hbar\omega_1}-e^{-\beta'_2\hbar\omega_2})}{\tau^{low}_R}.
\end{equation}

\par Our main result is the thermodynamical equivalence of Stirling cycles in the quantum regime of small temperature. Introducing appropriate temperature scales for the working system, we have shown that when it is small compared to all relevant energy scales of the system, both cycle types become equivalent. This equivalence emerges because, for small temperature, population of both the oscillators become indistinguishable. Remarkably, the equivalence also holds for overall cycle period and heat transfer rates. This is an artefact of the fact that at low temperatures, most of the population of harmonic oscillator working medium lies in the first two levels so that the bosonic oscillator is behaving like a two-level fermionic oscillator.

\section{CONCLUSION}\label{Sec. VII}

In this work we have presented a statistical generalization of bosonic and fermionic Stirling cycles with regenerative characteristics. The approach is based on oscillator models of working fluid that represents two distinct types of quantum Stirling cycles, with bounded and unbounded Hamiltonians of fermionic and bosonic statistics. The advantage of the proposed oscillator scheme is to provide a unified framework where both kinds of working systems can be depicted by the same set of physical parameters. This enables a meaningful comparison between two distinct types of statistical cycles, thus constitutes an essential ingredient of our theory. We now summarize our major conclusions as follows:

i) Unique generalization of heat and work in terms of change in frequency and population of the bosonic and fermionic oscillators can serve as a universal paradigm for the generalized version of first law of thermodynamics valid for both types of statistics.

ii) Apart from Newtonian law of heat conduction obeyed by harmonic oscillator cycle at high temperature, near equilibrium heat transfer rates between the fermionic and bosonic working systems and the heat reservoirs, exhibit in general a linear law of irreversible thermodynamics. While it is independent of the properties of the bath in the classical limit for both types of working fluids, their generic as well as the low temperature heat transfer coefficients strongly depend upon the particular choice of reservoir models.

iii) Reversible and irreversible performance of both the Stirling cycles become thermodynamically equivalent in the quantum limit. Equivalence also holds for low temperature heat transfer rates and the behaviour of cycle times. Validity regime of the equivalent performance is expressed in terms of energy and temperature scales of the working medium. Beyond this low temperature ``quantum'' limit, two models differ significantly, however, the nature of their maximum power behavior is analogous at both high and low temperature limits.

iv) Although the low temperature equivalence of harmonic and fermionic engine (refrigerator) are explicitly obtained for regenerative Stirling cycles, our general conclusions are expected to hold for other engine (refrigerator) cycles as well, since any nondegenerate multilevel system reduces to a two state fermionic oscillator at very low temperature.

\subsection*{ACKNOWLEDGEMENTS}

N. G. and S. B. thank respectively CSIR (JRF) and DST INSPIRE for fellowships. A. G. thanks Initiation Grant, IITK (IITK/CHM/2018513) and SRG (SERB/CHM/2019303), SERB, India for partial financial support.


%

\newpage
\onecolumngrid

\begin{center}
\textbf{APPENDIX}
\end{center}     
      
\setcounter{equation}{0}
    \renewcommand{\theequation}{A\arabic{equation}}
    \subsection{Derivation of the heat exchange}\label{Appendix-A}
    
From the population Eq.~\eqref{population} of our working medium we can write   
\begin{equation}
    \hbar\omega=K_BT\ln\left({\frac{1 \mp n}{n}}\right).
\end{equation}    
Using Eq.~\eqref{change-in-heat}, the amount of heat exchange at constant temperature $T$ (i.e. isothermal process) can be calculated as   
\begin{subequations}
\begin{eqnarray}
    Q_{i \rightarrow f}&=&\int_{n_i}^{n_f}\hbar\omega dn={K_BT}\int_{n_i}^{n_f}\ln\left({\frac{1 \mp n}{n}}\right)dn,\\
    &=&K_BT\left[{\int_{n_i}^{n_f}\ln(1 \mp n)dn-\int_{n_i}^{n_f}\ln(n)dn}\right],\\
    &=&K_BT\left[{n\ln\left(\frac{1 \mp n}{n}\right) \mp \ln(1 \mp n)}\right]\Bigg|^{n_f}_{n_i},\\
    &=&K_BT\left[{n_f\ln\left(\frac{1 \mp n_f}{n_f}\right)-n_i\ln\left(\frac{1 \mp n_i}{n_i}\right)\pm \ln\left(\frac{1 \mp n_i}{1 \mp n_f}\right)}\right],\\
    &=&\frac{\hbar\omega_{f}}{e^{\frac{\hbar\omega_f}{K_BT}} \pm 1}-\frac{\hbar\omega_{i}}{e^{\frac{\hbar\omega_i}{K_BT}} \pm 1} \pm K_B T\ln\left[{\frac{1 \pm e^{-\frac{\hbar\omega_{f}}{K_BT}}}{1 \pm e^{-\frac{\hbar\omega_{i}}{K_BT}}}}\right],
\end{eqnarray}
\end{subequations}
where $\omega_{i}$, $\omega_{f}$ are respectively the frequencies of the initial and final states. Similarly we can calculate the amount of heat exchange at constant frequency $\omega$ as  
\begin{subequations}
  	\begin{eqnarray}
  	  Q_{i \rightarrow f}&=&\hbar\omega\int_{n_i}^{n_f}dn=\hbar\omega(n_f-n_i),\\
  	&=&\hbar\omega\left(\frac{1}{e^{\frac{\hbar\omega}{K_BT_f}}\pm 1}-\frac{1}{e^{\frac{\hbar\omega}{K_BT_i}}\pm 1}\right),  
  	\end{eqnarray}
\end{subequations}
where $T_i$ and $T_f$ are respectively the initial and final temperatures for the process $i \rightarrow f$.

\end{document}